\documentclass[psfig,twocolumn,showpacs,preprintnumbers,amsmath,amssymb]{revtex4}
\usepackage{graphics}
\usepackage{dcolumn}
\usepackage{bm}

\begin{document}

\title{Dynamics of Boolean Networks with Scale-Free Topology}

\author{Maximino Aldana}
\email{maximino@control.uchicago.edu}
\affiliation{James Franck Institute, The University of Chicago.
  5640 South Ellis Avenue, Chicago, Il 60637, USA}

\date{\today}

\begin{abstract}
The dynamics of Boolean networks (the $N$-$K$ model) with scale-free
topology are studied here. The existence of a phase transition
governed by the value of the scale-free exponent of the network is
shown analytically by analyzing the overlap between two distinct
trajectories.  The phase diagram shows that the phase transition
occurs for values of the scale-free exponent in the open interval
$(2,2.5)$.  Since the Boolean networks under study are directed
graphs, the scale-free topology of the input connections and that of
the output connections are studied separately. Ultimately these two
topologies are shown to be equivalent. An important result of this
work is that the fine-tuning usually required to achieve stability in
Boolean networks with a totally random topology is no longer necessary
when the network topology is scale-free.

\end{abstract}

\maketitle

The study and characterization of the statistical properties of
complex networks has received renewed attention in the last few
years~\cite{01.4,01.5}. In particular, it has been recently shown by
A.  L. Barab\'asi, M. Newman, R. Sol\'e and many other authors that a
great variety of real networks exhibit a scale-free topology,
including the WWW and the Internet~\cite{web3,web1}, social
networks~\cite{collaboration1}, metabolic and protein
networks~\cite{metabolic1,metabolic2,metabolic3}, ecological
networks~\cite{ecology1,ecology2}, and genetic
networks~\cite{nk-scalefree1,nk-scalefree2}, to mention just a few
examples (for more references see~\cite{01.5,02.2}). A scale-free
topology means that the probability $P(k)$ that an arbitrary element
of the network is connected to exactly $k$ other elements has the form
$P(k)=Ck^{-\gamma}$, where $\gamma$ is usually called the scale-free
exponent. Scale-free networks have the key property that a small
fraction of the elements are highly connected whereas the majority of
the elements are poorly connected. The ubiquity of scale-free networks
has led to a systematic study of the structural properties that
characterize the wiring diagram of the network.  Nevertheless, the
dynamics generated by a scale-free network topology when the elements
are provided with some kind of dynamic interaction rule, remain
essentially unexplored.

An interesting dynamical network in which the scale-free topology has
important implications is the $N$-$K$ model proposed by Stuart
Kauffman in 1969 to describe generically the dynamics involved in the
processes of gene regulation and cell differentiation~\cite{69.1}. In
this classic model, the genome of a given organism is represented by a
set of $N$ genes, each being a binary variable describing the two
possible states of gene-expression: either the gene is expressed (1)
or it is not (0).  Since the expression of a gene is controlled by the
expression of some other genes, Kauffman assumed the genome as a
directed network in which a link from a given gene $A$ to another gene
$B$ means that $A$ controls the expression of $B$. In view of the
complexity exhibited by real genetic networks, Kauffman made three
simplifying assumptions: (a) every gene is connected (is controlled)
by exactly $K$ other genes; (b) the $K$ genes to which every gene is
connected are chosen randomly with uniform probability from the entire
system; (c) each gene is expressed with probability $p$ and is not
expressed with probability $1-p$, depending upon the configurations of
its $K$ controlling genes.

Even with these simplifying assumptions, a very rich and unexpected
behavior of the model was found (for references see \cite{02.3}). In
particular, in 1986 Derrida and Pomeau showed analytically the
existence of a dynamical phase transition controlled by the parameters
$K$ and $p$~\cite{86.3}. For every value of $p$ there exists a
critical value of the connectivity, $K_c(p)=[2p(1-p)]^{-1}$, such that
if $K<K_c(p)$ all perturbations in an initial state of the system die
out (ordered phase). For $K > K_c(p)$ a small perturbation in the
initial state of the system propagates across the entire system over
time (chaotic phase). According to Kauffman, Stauffer, and other
authors, only when $K=K_c(p)$ (the critical phase) does the $N$-$K$
model have the required stability properties compatible with the order
manifest in the genetic networks of living organisms
\cite{90.3,94.3}. This fact made Kauffman coin the term ``life at
the edge of chaos''.

\begin{figure}
\scalebox{0.5}{\includegraphics{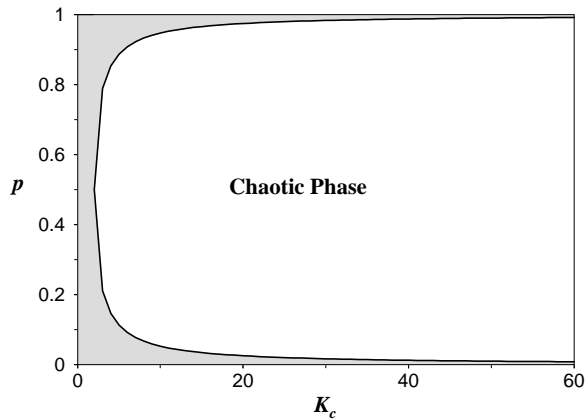}}
\caption{
\label{fig:nk-standard} 
Phase diagram for the standard $N$-$K$ model. The solid curve is the
critical connectivity $K_c$ as a function of $p$. The shaded area
represents the ordered phase ($K<K_c$) whereas the white area
represents the chaotic phase ($K>K_c$).}
\end{figure}

Although the $N$-$K$ model qualitatively points in the right
direction, it fails to account for a quantitative description of what
actually is observed in genetic networks. One of the main problems is
that the critical connectivity $K_c(p)$ is very small for most values
of $p$ (see Fig.~\ref{fig:nk-standard}). In contrast, it is well
known that real genetic networks exhibit a wide range of
connectivities.  For example, the expression of the human
$\beta$-globine gene (expressed in red blood cells) or the
\emph{even-skipped} gene in \emph{Drosophila} (playing an important
role in the development of the embryo), are each controlled by more
than 20 different regulatory proteins \cite{94.4}. Analogously, both
the the fibroblast growth factor (FGF) activation and the
platelet-derived growth factor (PDGF) activation in mammalian cells,
result in the cascade activation of over 60 other genes
\cite{00.4}. On the other side of the spectrum is the \emph{lac}
operon in \emph{E. coli}, which is regulated by only two proteins: the
\emph{lac} repressor protein and the catabolite activator protein
\cite{94.4}. These observations are contrary to the Kauffman model
since high connectivities imply either chaotic behavior or almost
constant boolean functions ($p$ very close to $0$ or $1$). To achieve
stability with moderate high connectivities in the $N$-$K$ model, it
is necessary to fine-tune the value of $p$. For instance, in order to
have ordered dynamics when the critical connectivity is $K_c=20$, the
value of $p$ should be in the interval $0<p<0.026$ (or
$0.974<p<1$). As far as we can tell, there is neither a theoretical
nor an experimental reason justifying why the parameters $K$ and $p$
should ``live'' in the shaded area of Fig.~\ref{fig:nk-standard}.

%
%
%
%

The above suggests that the random network topology upon which the
Kauffman model is based, is inadequate. In view of the ubiquity of
scale-free networks and of the fact that some genes in real genetic
networks are highly connected whereas others are not, it is reasonable
to replace assumptions (a) and (b) mentioned above by the assumption
that the connectivity $k$ of every gene in the network follows a
scale-free distribution, $P(k)\sim k^{-\gamma}$. It was not until very
recently that the dynamics of the $N$-$K$ model with a scale-free
topology were studied numerically
\cite{nk-scalefree1,nk-scalefree2}. Nevertheless, the values of
$\gamma$ and $p$ at which the phase transition occurs, if it does
occur at all, were unknown in these previous works and the results may
have to be reinterpreted.  In this letter we show analytically that
the $N$-$K$ model with a scale-free topology undergoes a phase
transition controlled by the scale-free exponent $\gamma$ and the
parameter $p$, and provide the phase diagram that fully identifies the
ordered, critical, and chaotic phases.

The model which we will be working with is the following. The network
is represented by a set of $N$ boolean variables (or elements),
$\{\sigma_1,\sigma_2,\dots,\sigma_N\}$. Each element $\sigma_i$ is
controlled by $k_i$ other elements of the network, where $k_i$ is
chosen randomly with probability $P_{\mathcal I}(k_i)$. Let
$\{\sigma_{i_1},\dots,\sigma_{i_{k_i}}\}$ be the set of the
controlling elements of $\sigma_i$. We then assign to each $\sigma_i$
a boolean function $f(\sigma_{i_1},\dots,\sigma_{i_{k_i}})$ such that
for each configuration of the controlling elements, $f_i=1$ with
probability $p$ and $f_i=0$ with probability $1-p$. Once the
controlling elements and the boolean functions are assigned to every
element in the network, the dynamics of the system is given by
$\sigma_i(t+1) = f_i(\sigma_{i_1}(t),\dots,\sigma_{i_{k_i}}(t))$.
We denote as $\Sigma_t$ the state of the entire system at time $t$,
$\Sigma_t=\{\sigma_1(t),\sigma_2(t),\dots,\sigma_N(t)\}$.

To show the existence of the phase transition we consider the overlap
$x(t)$ between two distinct configurations $\Sigma_t$ and
$\tilde{\Sigma}_t$ (the temporal evolution of these two configurations
is governed by the same set of boolean functions). The overlap is
defined as the fraction of elements in both configurations that have
the same value, and is given by
\begin{equation}
x(t) = 1-\frac{1}{N}\sum_{i=1}^N|\sigma_i(t)-\tilde{\sigma}_i(t)|.
\label{overlap}
\end{equation}
If $\Sigma_t$ and $\tilde{\Sigma}_t$ are totally independent, then
$x(t)\approx0.5$ whereas if they are almost equal $x(t)\approx1$. In
the limit $N\rightarrow\infty$ the overlap becomes the probability for
two arbitrary but corresponding elements, $\sigma_i(t)\in\Sigma_t$ and
$\tilde{\sigma}_i(t)\in\tilde{\Sigma}_t$, to be equal.  The stationary
value of the overlap, defined as $x=\lim_{t\rightarrow\infty} x(t)$,
can be considered as an order parameter of the system. If $x=1$, the
system is insensitive to initial perturbations (all differences
between configurations die out over time). In this case the system
presents an ordered behavior.  On the other hand, if $x\neq1$, the
initial perturbations propagate across the entire system and do not
disappear. In this case the system exhibits a chaotic behavior.

%
%
\begin{figure}
\scalebox{0.5}{\includegraphics{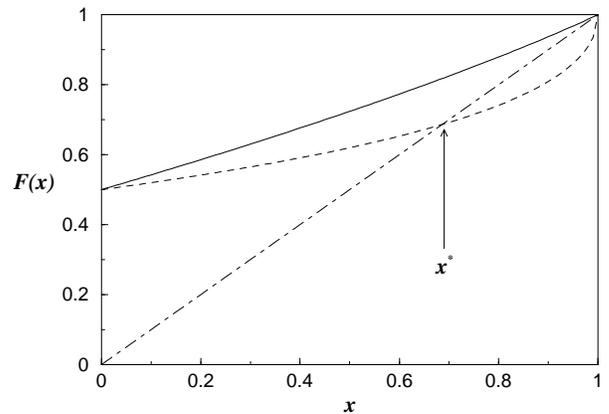}}
\caption{
\label{fig:mapping} 
Plot of the mapping $F(x)$ as a function of $x$ for a scale-free
distribution $P_{\mathcal I}(k)$. The solid curve corresponds to
$\gamma=3$, for which $\lim_{x\rightarrow1^-}dF(x)/dx < 1$. The only
fixed point in this case is $x=1$. The dashed curve corresponds to
$\gamma=1.5$, for which $\lim_{x\rightarrow1^-}dF(x)/dx =\infty$.  In
this other case there is a fixed point $x^*\neq1$. The dashed-dotted
line is the identity.}
\end{figure}
%
%
%

By generalizing the annealed computation carried out by Derrida and
Pomeau in Ref.~\cite{86.3} to the case in which each element receives
$k$ inputs with probability $P_{\mathcal I}(k)$, one finds that the
overlap obeys the dynamical equation
\begin{subequations}
\label{eq:mapping}
\begin{equation}
x(t+1) = F\left(x(t)\right),
\label{eq:fixedpoint}
\end{equation}
where the mapping $F(x)$ is given by
\begin{equation}
F(x) \equiv 1-2p(1-p)\Big\{1- \sum_{k=1}^\infty x^k 
P_{\mathcal I}(k)\Big\}.
\label{eq:F}
\end{equation}
\end{subequations}
In the limit $t\rightarrow\infty$, Eq.~(\ref{eq:mapping}) becomes
the fixed point equation $x=F(x)$ for the stationary value of the
overlap. Note that $x=1$ is always a fixed point of
Eq.~(\ref{eq:mapping}).  Nonetheless, this solution may be stable
or unstable depending on $P_{\mathcal I}(k)$. Note also that $F(x)$ is
a monotonically increasing function of $x$ with the property that
$F(0)=1-2p(1-p)$ and $F(1)=1$. Therefore, Eq.~(\ref{eq:mapping})
will have a stable fixed point $x^*\neq1$ only if
$\lim_{x\rightarrow1^-}dF(x)/dx > 1$ (chaotic phase). In contrast, if
$\lim_{x\rightarrow1^-}dF(x)/dx < 1$ the only fixed point is $x=1$
(ordered phase). The situation is illustrated in
Fig.~\ref{fig:mapping}. The phase transition between the ordered and
chaotic regimes occurs when $\lim_{x\rightarrow1^-}dF(x)/dx = 1$. From
Eq.~(\ref{eq:F}) it follows that
\begin{eqnarray}
\lim_{x\rightarrow1^-}\frac{dF(x)}{dx}&=&2p(1-p)
\sum_{k=1}^\infty kP_{\mathcal I}(k) \nonumber \\
&=& 2p(1-p)\langle k\rangle_{\mathcal I},
\label{eq:F-slope}
\end{eqnarray}
where $\langle k\rangle_{\mathcal I}=\sum_{k=1}^\infty kP_{\mathcal
I}(k)$ is the first moment of $P_{\mathcal I}(k)$. The phase
transition is then determined by the condition
\begin{equation} 
2p(1-p)\langle k\rangle_{\mathcal I} = 1.
\label{eq:phasetransition}
\end{equation}
In the standard $N$-$K$ model all the elements have the same
connectivity, $k_i=\langle k \rangle_{\mathcal I} = K$, and
Eq.~(\ref{eq:phasetransition}) reduces to the result obtained by
Derrida and Pomeau in Ref.~\cite{86.3}.  (Luque and Sol\'e derived
Eq.~(\ref{eq:phasetransition}) in Ref.~\cite{97.9} by considering the
set of relevant elements of the system---the elements that do not
reach a constant value over time.)

It is interesting to note that the phase transition is governed only
by the first moment of $P_{\mathcal I}(t)$. Nevertheless, for the
scale-free distribution $P_{\mathcal I}(k)=Ck^{-\gamma}$, the first
moment is not necessarily a meaningful parameter to characterize the
network topology. For instance, if $2 < \gamma \leq 3$, the second
moment of the distribution is infinite even when the first moment has
a finite value, which means that the fluctuations around the first
moment are much larger than the first moment itself. Therefore, rather
than characterizing the phase transition using the first moment of the
distribution, we will do it by means of the scale-free exponent
$\gamma$, which is the only natural parameter that determines
the network topology.

For the scale-free distribution to be normalizable, it is
necessary to have $\gamma > 1$. Under such conditions, the
probability function $P_{\mathcal I}(k)$ is given by
\begin{equation}
P_{\mathcal I}(k) = \frac{1}{\zeta(\gamma)} k^{-\gamma},
\label{eq:distribution}
\end{equation}
where $\zeta(\gamma)=\sum_{k=1}^\infty k^{-\gamma}$ is the Riemann
Zeta function. The first moment of this distribution is then expressed
as a function of $\gamma$ as,
\begin{equation}
\langle k\rangle_{\mathcal I} = \left\{
\begin{array}{lll}
\infty & & \mbox{if } 1< \gamma \leq 2 \\
 & & \\
\frac{\zeta(\gamma-1)}{\zeta(\gamma)} & & \mbox{if } \gamma > 2
\end{array}
\right. .
\label{eq:firstmoment}
\end{equation}
From Eqs.~(\ref{eq:F-slope})--(\ref{eq:firstmoment}), it follows that
the fixed point $x=1$ is unstable if $1<\gamma\leq2$. In this case
Eq.~(\ref{eq:mapping}) has a stable fixed point $x^*\neq1$, and
the system is in the chaotic phase for any value of $p$ in the open
interval $(0,1)$.  The only way in which the overlap between two
distinct configurations can converge to $1$ is if all the boolean
functions are constant, namely, if either $p=0$ or $p=1$. On the other
hand, when $\gamma>2$ the first moment of the distribution is
finite. In this case, the value $\gamma_c$ of the scaling-free
exponent at which the phase transition occurs is determined by the
transcendental equation,
\begin{equation}
2p(1-p)\frac{\zeta(\gamma_c-1)}{\zeta(\gamma_c)}=1.
\label{eq:phasediagram}
\end{equation}
%

%
\begin{figure}
\scalebox{0.5}{\includegraphics{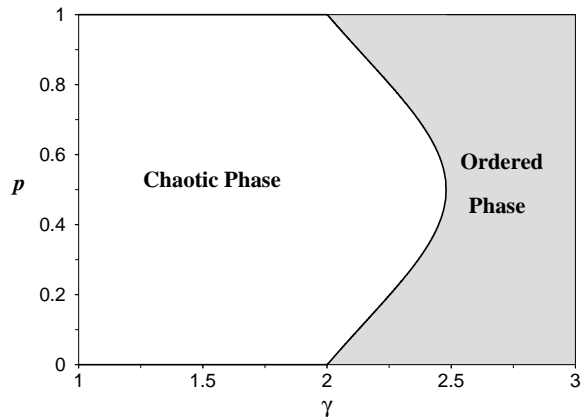}}
\caption{
\label{fig:nk-scalefree} 
Phase diagram for the $N$-$K$ model with a scale-free network
topology.  The solid line is the value of $\gamma_c$ as a function of
$p$ given by the transcendental equation (\ref{eq:phasediagram}). The
white and shaded areas represent the chaotic phase
($\gamma<\gamma_c$) and the ordered phase ($\gamma>\gamma_c$),
respectively.}
\end{figure}
%
%

The values $\gamma_c$ and $p$ for which this equation is satisfied are
plotted in Fig.~\ref{fig:nk-scalefree}. As can be seen,
$\gamma_c\in[2,2.5]$ for any value of $p$. Actually, $\gamma_c$
reaches its maximum value $\gamma_c^{max}\approx2.47875$ at
$p=0.5$. Above this maximum value the system is always in the ordered
phase regardless of the value or $p$. It is interesting to note that
$\gamma\in[2,3]$ for the majority of the real scale-free networks that
have been analyzed \cite{01.5}.

An important characteristic of the $N$-$K$ model is that it is a
directed graph: if $\sigma_i$ is an input of $\sigma_j$, the opposite
does not necessarily occur.  Every element $\sigma_i$ is regulated by
$k_i$ elements. But $\sigma_i$ can in turn regulate the value of a
number of other elements, say $l_i$. The distribution $P_{\mathcal
I}(k)$ of input connections is not necessarily equal to the
distribution $P_{o}(l)$ of output connections. Up to now we have
assumed that $P_{\mathcal I}(k)$ is a scale-free distribution. But as
the example of the fibroblast growth factor (FGF) suggests, it is also
possible for the network of output connections to present a scale-free
topology. To apply our previous results to the case in which
$P_{o}(l)$ is known instead of $P_{\mathcal I}(k)$, we should find how
these two distributions are related.

Suppose that an arbitrary element $\sigma_i$ of the network has $l_i$
outputs with probability $P_{o}(l_i) =
l_i^{-\gamma}/\zeta(\gamma)$. (We assume that $\gamma>2$ so that the
first moment of the distribution is well defined.) When the $l_i$
outputs of every element are chosen randomly with uniform probability
from the entire system, the input probability distribution
$P_{\mathcal I}(k)$ is given by
\begin{equation}
P_{\mathcal I}(k)=\binom N K \left(\frac{\langle
l\rangle_{o}}{N}\right)^k
\left(1-\frac{\langle l\rangle_{o}}{N}\right)^{N-k},
\label{eq:binomial}
\end{equation}
where $\langle l\rangle_{o}$ is the first moment of $P_{o}(l)$.  In
the limit $N\rightarrow\infty$, the above expression transforms into
the Poisson distribution:
\begin{equation}
P_{\mathcal I}(k)= e^{-\langle l\rangle_{o}}
\frac{\left[\langle l\rangle_{o}\right]^k}{k!}.
\label{eq:poisson}
\end{equation}
This result shows that the fist moment of the input distribution
$P_{\mathcal I}(k)$ is equal to the first moment of the output
distribution $P_{o}(k)$, $\langle k\rangle_{\mathcal I} =
\langle l\rangle_{o}$. Therefore, Eq.~(\ref{eq:phasetransition}) and
the phase diagram shown in Fig.~\ref{fig:nk-scalefree} are still valid
if we substitute $\langle l\rangle_{\mathcal I}$ by $\langle
l\rangle_{o}$.

A consequence of the scale-free network topology is that the
constraint of having a very low connectivity for every element (or a
high connectivity with almost constant functions) in order to be in
the ordered phase, is no longer valid. The scale-free topology
eliminates the necessity of fine-tuning the parameters $K$ and $p$ to
achieve stable dynamics. Furthermore, the fact that the phase
transition occurs for values of $\gamma$ in the interval $(2,2.5)$,
allows the existence of elements with a wide range of connectivities
in the critical and ordered phases, which is required to describe the
observed behavior of real genetic networks. Now that the human genome,
as well as the genome of some other organisms, have been thoroughly
sequenced, it will be possible to determine experimentally if the
topology of real genetic networks is scale-free or
not \footnote{Recent preliminary results indicate that the
gene-regulatory network of
\emph{E. coli}, and the molecular interaction map of the mammalian
cell cycle control, have scale-free topologies
\cite{nk-scalefree1,nk-scalefree2}.  However, these studies were
carried out with partial information of the genetic networks and
under the assumption that the important parameter characterizing
the network topology is the mean connectivity rather than the
scale-free exponent.}.  
More work is called for to fully characterize
the statistical properties of Boolean networks with scale-free
topology, such as the distribution of the number of different orbits,
the distribution of the orbit lengths, and the stability of the
dynamics in the three different phases of the $N$-$K$ model. We
believe that the phase diagram shown in Fig.~\ref{fig:nk-scalefree}
will be a useful guide for further studies.

I would like to thank Leo P. Kadanoff and Philippe Cluzel for useful
and enlightening discussions. I also thank Leo Silbert and Nate Bode
for their comments and suggestions. This work was supported in part by
the MRSEC Program of the NSF under award number 9808595, and by the
NSF DMR 0094569. M. Aldana also acknowledges the Santa Fe Institute of
Complex Systems for partial support through the David and Lucile
Packard Foundation Program in the Study of Robustness.


\end{document}